\documentclass[twocolumn]{aastex631}
\usepackage{apjfonts}
\usepackage{graphicx}
\usepackage{amsmath}
\usepackage{amssymb}
\usepackage{color}

\gdef\xx[#1]{\textcolor{red}{#1}}

\gdef\dft{DF2}
\gdef\dff{DF4}

\newcommand{\GG}[1]{}
\lefthead{van Dokkum et al.}
\righthead{}
\begin{document}

\newcommand\XXX[1]{{\textcolor{red}{\textbf{x\ #1\ x}}}}

\title{Monochromatic globular clusters as a critical test of formation
models for the dark matter deficient galaxies NGC\,1052-DF2 and NGC\,1052-DF4}


\author[0000-0002-8282-9888]{Pieter van Dokkum}
\affiliation{Astronomy Department, Yale University, 52 Hillhouse Ave,
New Haven, CT 06511, USA}
\author[0000-0002-5120-1684]{Zili Shen}
\affiliation{Astronomy Department, Yale University, 52 Hillhouse Ave,
New Haven, CT 06511, USA}
\author[0000-0003-2473-0369]{Aaron J.\ Romanowsky}
\affiliation{Department of Physics and Astronomy, San Jos\'e State University,
San Jose, CA 95192, USA}
\affiliation{Department of
Astronomy and Astrophysics, University of California Santa Cruz, 1156 High Street, Santa Cruz, CA 95064, USA}
\author[0000-0002-4542-921X]{Roberto Abraham}
\affiliation{Department of Astronomy \& Astrophysics, University of Toronto,
50 St.\ George Street, Toronto, ON M5S 3H4, Canada}
\author[0000-0002-1590-8551]{Charlie Conroy}
\affiliation{Harvard-Smithsonian Center for Astrophysics, 60 Garden Street,
Cambridge, MA, USA}
\author[0000-0002-1841-2252]{Shany Danieli}
\altaffiliation{NASA Hubble Fellow}
\affiliation{Department of Astrophysical Sciences, 4 Ivy Lane, Princeton University, Princeton, NJ 08544, USA}
\author[0000-0003-0250-3827]{Dhruba Dutta Chowdhury}
\affiliation{Astronomy Department, Yale University, 52 Hillhouse Ave,
New Haven, CT 06511, USA}
\author[0000-0002-7743-2501]{Michael A.\ Keim}
\affiliation{Astronomy Department, Yale University, 52 Hillhouse Ave,
New Haven, CT 06511, USA}
\author[0000-0002-8804-0212]{J.~M.\ Diederik Kruijssen}
\affiliation{Astronomisches Rechen-Institut, Zentrum f\"ur Astronomie der Universit\"at
Heidelberg, M\"onchhofstra\ss{}e 12--14, D-69120 Heidelberg, Germany}
\author[0000-0001-6755-1315]{Joel Leja}
\affiliation{Department of Astronomy \& Astrophysics, Penn State University,
525 Davey Laboratory, University Park, PA 16802}
\author[0000-0003-2482-0049]{Sebastian Trujillo-Gomez}
\affiliation{Astronomisches Rechen-Institut, Zentrum f\"ur Astronomie der Universit\"at
Heidelberg, M\"onchhofstra\ss{}e 12--14, D-69120 Heidelberg, Germany}

\begin{abstract}

It was recently proposed that the dark matter-deficient ultra-diffuse galaxies
DF2 and DF4 in the NGC\,1052 group could be the
products of a ``bullet dwarf'' collision between two gas-rich progenitor galaxies.
In this model \dft\ and \dff\ formed at the same time in the immediate aftermath of the collision, and a strong prediction is that their globular clusters should have nearly identical stellar populations.
Here we test this prediction by measuring accurate $V_{606}-I_{814}$
colors from deep HST/ACS imaging. We find that the clusters are extremely homogeneous.
The mean color difference between the globular clusters in \dft\ and \dff\ is
$\Delta_{\rm DF2-DF4} = -0.003\pm 0.005$\,mag and
the observed
scatter for the combined sample of 18 clusters with $M_{606}<-8.6$ in both galaxies is
$\sigma_{\rm obs} = 0.015 \pm 0.002$\,mag.  After accounting
for observational uncertainties and
stochastic cluster-to-cluster variation in the number of red giants,
the remaining scatter is $\sigma_{\rm intr}=0.008^{+0.005}_{-0.006}$\,mag. Both the
color difference and the scatter are an order of
magnitude smaller than in other dwarf galaxies, and we infer that the
bullet scenario passes an important test that could have falsified it.
No other formation models have predicted this extreme uniformity of the globular clusters in the two galaxies.
We find that the galaxies themselves are slightly
redder than the clusters, consistent with a previously-measured metallicity difference.
Numerical simulations have shown that
such differences are expected in the bullet scenario, as the
galaxies continued to self-enrich after the formation of the globular clusters.

\end{abstract}


\section{Introduction}

NGC\,1052-DF2 (or \dft) and NGC\,1052-DF4 (\dff) share two unusual properties that set them apart
from all other known galaxies. First, their globular clusters are,
on average, a factor of $\approx 4$ brighter and a factor of $\approx 2$ larger than canonical
values \citep{dokkum:18b,dokkum:19df4,ma:20,shen:21a}. 
Furthermore, their
velocity dispersions are consistent with their stellar mass alone and much
smaller than expected from a normal dark matter halo \citep{dokkum:18,wasserman:18,danieli:19,emsellem:19}.

Initially the main question was whether the dynamical masses and distances
were measured correctly \citep[see, e.g.,][]{martin:18,laporte:19,trujillo:19}, but as  the anomalous properties 
of the galaxies were gradually confirmed \citep[and corroborated with independent evidence; see][]{dutta:19,keim:21}
the focus shifted to the question how they were formed.
Proposals include assembly out of tidally-removed gas
\citep{fensch:19}, stripping of dark matter by close encounters
with NGC\,1052 \citep{ogiya:18,carleton:19,nusser:20,ogiya:22,moreno:22}
or NGC\,1035 \citep{montes:20},
jet- or outflow-induced star formation, like Minkowski's object \citep{vanbreugel:85,natarajan:98}, 
and extreme feedback in low mass halos \citep{trujillogomez:22}.

\citet{silk:19} suggested that DF2 could be the result of a
``mini bullet-cluster'' event, where the dark matter and the baryons became separated in a nearly head-on
encounter between two gas-rich progenitor galaxies. Such a collision produces two dark matter-dominated
remnants with a globular cluster-rich, dark matter-free object in between them that formed from the shocked gas. 
This scenario was further explored with simulations in \citet{shin:20} and \citet{lee:21}, who showed
that collisions between an unbound object and a satellite could explain many of the observed properties
of the galaxies and occur with some regularity in cosmological simulations. The main issue with this
model -- as with many of the alternative explanations listed above -- is the presence of {\em two}
dark matter-deficient galaxies, seemingly requiring lightning to strike twice in the same group.

Recently we suggested that a single bullet
dwarf collision may have produced {\em both} \dft\ and \dff\ \citep{dokkum:22}.
This joint formation is consistent with the striking similarities between the two galaxies,
their radial velocities and line-of-sight distances,  the emergence of
multiple clumps in at least some bullet collision simulations \citep{shin:20},
and with
the discovery that \dft\ and \dff\ are part of a remarkable trail of $\approx 10$ galaxies in the group.

As noted in \citet{dokkum:22}
this formation model for DF2 and DF4 is falsifiable, as it makes a very specific prediction for their
globular clusters. Both galaxies formed out of the gas that was left behind by the progenitor
galaxies, at the same time. This gas mixed efficiently during the collision and should have a uniform
metallicity.
While overall star formation likely lasted for $\gtrsim 500$\,Myr \citep{shin:20,lee:21}, 
the globular clusters formed almost instantaneously before further enrichment could substantially
change the abundance of the gas \citep{lee:21}.
As a result, the stellar populations of all the
globular clusters, in both galaxies, should be {\em extremely}
similar.\footnote{We note here that simulations have so far only focused on comparing the properties of clusters across a single resulting galaxy fragment, and not yet across several galaxies.} This is not what is typically observed
in dwarf galaxies. In a comprehensive study of the colors of globular clusters in Virgo dwarf galaxies,
\citet{peng:06} found that there is substantial cluster-to-cluster
and galaxy-to-galaxy scatter.

Previous measurements of the colors of the globular clusters in DF2 and DF4 do in fact suggest differences between
them, in potential conflict with the bullet model. Tables 1 and 2 in \cite{shen:21a} 
imply a mean $V_{606}-I_{814}$ globular cluster color 
of $0.37\pm 0.01$ AB mag in DF2 and $0.41\pm 0.01$ in
DF4.\footnote{{Throughout, $V_{606}-I_{814}$ is used to denote the ${\rm F606W} - {\rm F814W}$ color in the AB system. The conversion to Johnson-Cousins $V-I$ in the Vega system is $V-I = 1.312(V_{606}-I_{814}) + 0.364$ \citep{sirianni:05}.}}
This difference seems small but it corresponds to $\approx 0.3$\,dex in age or $\approx 0.4$\,dex in metallicity.
Furthermore, the cluster-to-cluster scatter is non-zero, with $\sigma= 0.04 \pm 0.01$\,mag
for both galaxies. There is also a hint that the galaxies themselves might have different $V_{606}-I_{814}$ colors:
\cite{cohen:18} finds $0.40$ for the diffuse light in DF2 and
$0.32$ for DF4, albeit with an uncertainty of $0.1$ for both.

These measurements were performed using standard techniques \citep[such as Source Extractor;][]{bertin:96} and
are largely based on single-orbit HST/ACS images that were reduced with the default STScI flat fields.
Here we remeasure the colors of the clusters and the
diffuse light in both DF2 and DF4 using custom techniques and well-calibrated, much deeper data,
as a stringent test of the bullet dwarf model.
Where needed we use a distance of $D=21.7$\,Mpc for
DF2 and $D=20.0$\,Mpc for DF4 \citep[][Z.~Shen et al., in preparation]{danieli:20,shen:21b}.

\begin{figure*}[htbp]
  \begin{center}
  \includegraphics[width=0.9\linewidth]{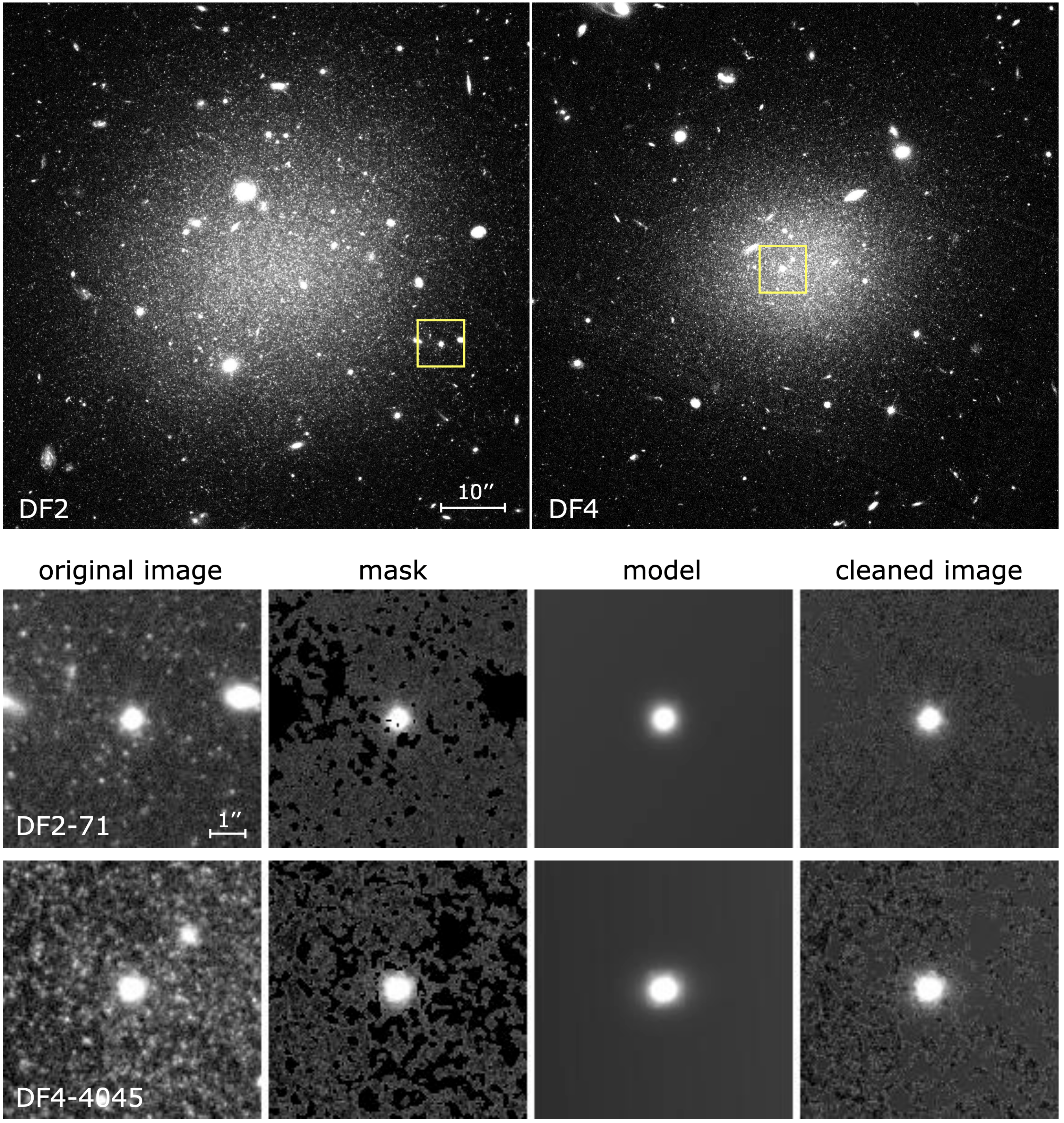}
  \end{center}
\vspace{-0.2cm}
    \caption{
{Top panels:} Images of DF2 and DF4 in the HST/ACS $I_{814}$ filter, after applying flat field corrections to the
{\tt flc} files and drizzling. The images span $80\arcsec \times 80\arcsec$. {Bottom panels:} Photometry procedure
for two of the globular clusters, indicated with yellow boxes above. From left to right we show
the original $I_{814}$ image (with a size of $7\arcsec \times 7\arcsec$), the final object mask, the PSF-convolved
\cite{king:62} model fit with best-fitting background plane, and the cleaned image. The cleaned image is
the original image with the masked areas filled in by the King model. Colors and total magnitudes are
measured from aperture photometry on the cleaned $V_{606}$ and $I_{814}$ images.
}
\label{figure1.fig}
\end{figure*}

\section{Data}

We make use of deep HST/ACS data that were obtained in programs GO-15695 and GO-15851.
The aim of these programs was to measure distances to DF4 (GO-15695) and DF2 (GO-15851) from the
tip of the red giant branch (TRGB). For DF4 the exposure
times were 7 and 3 orbits in $I_{814}$ and $V_{606}$ respectively.
In \cite{danieli:20} these were combined with the $1+1$ orbits that had been obtained in GO-14644
\citep[see][]{cohen:18}, and the location of the TRGB was measured from these $8+4$ orbit depth data.
The DF2 observations of GO-15851 were deeper, at 19 orbits in $I_{814}$ and 19 orbits in $V_{606}$.
\citet{shen:21b} used these data to measure the TRGB distance of DF2, again after adding the GO-14644 data
for a total exposure of $20+20$ orbits.

In this study we redrizzle the ACS data, with several changes. First, we discard the $1+1$ orbits
that were obtained for DF2 and DF4 in GO-14644. The orientation of these observations was very different
from the more recent data, which means the point spread function (PSF)
and charge-transfer efficiency corrections are different, and
there is no need to maximize depth:
the fluxes of the globular clusters are typically $\sim 30$\,e$^{-}$\,s$^{-1}$, which means that
Poisson errors are $\lesssim 0.5$\,\% even in a single orbit. 

More importantly, we apply a flat-fielding correction to the {\tt flc} files prior to drizzling.
The ACS flat fields were last updated in 2006 and they no longer provide optimal corrections.
As described in instrument science report\footnote{The repository of these reports is \url{https://www.stsci.edu/hst/instrumentation/acs/documentation/instrument-science-reports-isrs}.} ACS\,2017-09,
ultra-deep stacks of Frontier Fields images show that there are systematic
flat-fielding residuals at the level of $1$\,\%. This effect is also described in ISR ACS\,2020-08.
The residuals correlate with the local thickness of the CCD, and as the correlation between quantum efficiency and
thickness reverses at $\approx 700$\,nm, the residuals in $V_{606}$ and $I_{814}$ are spatially anti-correlated.
As a result,  flat-fielding errors in the $V_{606}-I_{814}$ color reach $2$\,\%, even though they
are only half that in each filter individually.

The ACS team at STScI provided us with preliminary correction flats and we applied these to the {\tt flc} files.
The {\tt flc} files were aligned with each other using {\tt tweakreg}. The {\tt tweakreg} algorithm is sensitive
to cosmic rays; we addressed this by running the code on
versions of the data where cosmic rays were removed \citep[with {\tt L.A.Cosmic}][]{dokkumc:01}.
$I_{814}$ images of DF2 and DF4 are shown in Fig.\ \ref{figure1.fig}. Visually
comparing the default and flat-field-corrected images, there is a clear improvement in the flatness
of the sky background, particularly for DF2. From the remaining variation in the background
we estimate that residual flat fielding errors are
$0.5\pm 0.2$\,\%. Assuming that the errors are no longer (anti-)correlated,  the uncertainty in the color due to spatially-varying
flat-fielding uncertainties is then $0.7\pm 0.3$\,\% in our images.

\section{Measurements of Colors and Magnitudes}

\subsection{Sample}

At present 24 globular clusters have been spectroscopically-confirmed, 12 in DF2 and 12 in DF4. These are
listed in Tables 1 and 2 of \citet{shen:21a}; the sources of the spectra are \citet{dokkum:18b},
\citet{dokkum:19df4}, \citet{emsellem:19}, and \citet{shen:21a}.
The sample is likely complete for $M_V\lesssim -8.5$. Several clusters are not included in our analysis.
DF2-80 is superposed on a faint spectroscopically-identified
blue background galaxy \citep{shen:21a} with no possibility
of disentangling the two objects; DF4-97 falls outside of the coverage of the GO-15695 data; and DF4-926 is
projected onto a bright background merger \citep[see][]{dokkum:19df4}. This leaves 11 confirmed clusters
in DF2 and 10 in DF4.

Three of the confirmed DF4 clusters are fainter than $M_V=-8$.
As we show later the errors increase sharply at fainter magnitudes, and our quantitative
analysis of the $V_{606}-I_{814}$ scatter
focuses on the $11+7=18$ objects that have $M_V<-8.6$. The faint clusters are used to investigate
trends with magnitude; to this end
we use an augmented sample in \S\,\ref{obs.sec} that includes five faint objects without a spectrum. These
candidate clusters satisfy
the \citet{shen:21a} color and size criteria and are located within $45\arcsec$ of the center of DF2 or DF4.
The sample is listed in Table 1.

\subsection{Cleaned Globular Cluster Images}

Even with flat fielding errors largely eliminated,
it is not straightforward to obtain photometry with the desired 
accuracy. An investigation into the effects of changes in aperture sizes and the background subtraction
shows seemingly random variation in the colors at the 1\%\,--\,2\% level for individual clusters.
These variations can be traced to the cumulative effects of contaminating objects, mostly red giants in the main
bodies of the galaxies, on both the flux measurements and the estimation of the background.

In shallow data this is an unavoidable source of noise but owing to
the depth of the $I_{814}$ images we can mitigate these effects. 
We use the following procedure for
each cluster. We begin by creating a mask for
the brightest contaminating objects in a $7\arcsec \times 7\arcsec$ region centered on the cluster.
{All pixels with a flux in excess of} $0.03$\,e$^{-}$\,s$^{-1}$\,pix$^{-1}$ {are masked}, excluding the central
$1\farcs 0 \times 1\farcs 0$ to avoid masking the cluster itself.
{The purpose of this initial mask is to enable an initial
fit to the globular cluster.}
A PSF-convolved 2D modified King profile
with $\alpha=2$ \citep{elson:99,galfit} is fit to the image, using
{\tt galfit} \citep{galfit}. The bright {pixel}
mask is used, the background is modeled as a constant,
and for the PSF we choose a non-saturated star in the image.

With a model for the cluster in hand an improved mask can be created by identifying pixels that deviate
significantly from the model. Specifically, we create a residual image $R=(I - M)/M_{0.02}$, with $I$
the image, $M$ the model, and $M_{0.02}$ the model with all pixels $\leq 0.02$\,e$^{-}$\,s$^{-1}$\,pix$^{-1}$
set to 0.02. After median filtering $R$ by a $3\times 3$ pixel filter, all pixels $>0.15$ are flagged
and added to the initial mask. This process masks objects whose median flux in a $3\times 3$ pixel
aperture deviates by more than 15\,\%
from the model, as well as objects away from the central regions whose flux exceeds a fixed threshold
of $\gtrsim 0.003$\,e$^{-}$\,s$^{-1}$ (with the precise threshold depending on the background level). {The threshold of
$0.02$\,e$^{-}$\,s$^{-1}$\,pix$^{-1}$ was chosen to ensure a smooth transition between these regimes.}
The chosen values are a compromise between masking as many contaminants
as possible and leaving sufficient data for reliable flux and background measurements.
The masks for two representative clusters, DF2-71 and DF4-4045, are
shown in Fig.\ \ref{figure1.fig}. 

The fit to the globular cluster is repeated with this new mask, now modeling the background as a plane
with gradients in $x$ and $y$ to properly account for the diffuse light of DF2 and DF4. The models
for DF2-71 and DF4-4045 are shown in the third column of Fig.\ \ref{figure1.fig}.
Finally, a cleaned image is created by replacing all masked pixels in the original image by the corresponding pixels
in the model (fourth column of Fig.\ \ref{figure1.fig}).

\begin{deluxetable}{lrcccc}
\tablecolumns{6}
\tablewidth{0pt}
\tablecaption{Photometry of globular clusters
\label{pho.tab}}
\tablehead{
  \colhead{Id} & \colhead{$M_{606}$} & \colhead{$V_{606}-I_{814}$}
  & \colhead{$\pm$\,meas} & \colhead{$\pm$\,phot} & \colhead{$\pm$\,stoch}}
\startdata
\multicolumn{6}{c}{\em Main sample:}\\
\hline
DF2-39 & $-9.40$ & 0.363 & 0.001 & 0.007 & 0.008 \\ 
DF2-59 & $-8.97$ & 0.389 & 0.004 & 0.008 & 0.010 \\ 
DF2-71 & $-9.12$ & 0.396 & 0.007 & 0.010 & 0.009 \\ 
DF2-73 & $-10.24$ & 0.365 & 0.002 & 0.007 & 0.007 \\ 
DF2-77 & $-9.76$ & 0.356 & 0.006 & 0.009 & 0.007 \\ 
DF2-85 & $-9.35$ & 0.388 & 0.010 & 0.012 & 0.008 \\ 
DF2-91 & $-9.29$ & 0.377 & 0.005 & 0.008 & 0.009 \\ 
DF2-92 & $-9.61$ & 0.382 & 0.005 & 0.009 & 0.007 \\ 
DF2-93 & $-8.89$ & 0.349 & 0.011 & 0.013 & 0.010 \\ 
DF2-98 & $-8.97$ & 0.374 & 0.003 & 0.008 & 0.010 \\ 
DF2-101 & $-8.88$ & 0.365 & 0.003 & 0.008 & 0.010 \\ 
DF4-1092 & $-9.12$ & 0.377 & 0.002 & 0.007 & 0.008 \\ 
DF4-1403 & $-8.96$ & 0.384 & 0.004 & 0.008 & 0.009 \\ 
DF4-3515 & $-8.67$ & 0.374 & 0.005 & 0.009 & 0.010 \\ 
DF4-4045 & $-9.69$ & 0.375 & 0.006 & 0.009 & 0.007 \\ 
DF4-5675 & $-9.29$ & 0.360 & 0.004 & 0.008 & 0.008 \\ 
DF4-6129 & $-9.13$ & 0.411 & 0.004 & 0.008 & 0.008 \\ 
DF4-6571 & $-8.93$ & 0.381 & 0.004 & 0.008 & 0.009 \\ 
\hline
\multicolumn{6}{c}{\em Faint sample:}\\
\hline
DF2-2764 & $-8.02$ & 0.417 & 0.026 & 0.027 & 0.016 \\ 
DF2-3555 & $-7.70$ & 0.391 & 0.017 & 0.018 & 0.017 \\ 
DF2-4061 & $-8.02$ & 0.378 & 0.020 & 0.021 & 0.016 \\ 
DF4-1755 & $-7.92$ & 0.429 & 0.015 & 0.017 & 0.016 \\ 
DF4-3278 & $-8.17$ & 0.324 & 0.026 & 0.027 & 0.013 \\ 
DF4-3602 & $-8.57$ & 0.427 & 0.021 & 0.022 & 0.011 \\ 
DF4-4176 & $-7.25$ & 0.379 & 0.070 & 0.071 & 0.023 \\ 
DF4-4366 & $-7.40$ & 0.429 & 0.018 & 0.019 & 0.020 \\ 
\enddata
\tablecomments{Magnitudes are on the AB system.
IDs are from \citet{shen:21a}. Uncertainties on the colors are
$\pm 1\sigma$, with $\pm$\,meas the measurement uncertainty,
$\pm$\,phot the measurement uncertainty combined with
the flat fielding uncertainty, and $\pm$\,stoch the stochastic
uncertainty due to variations in the number of red giants.}
\end{deluxetable}

\begin{figure*}[htbp]
  \begin{center}
  \includegraphics[width=0.9\linewidth]{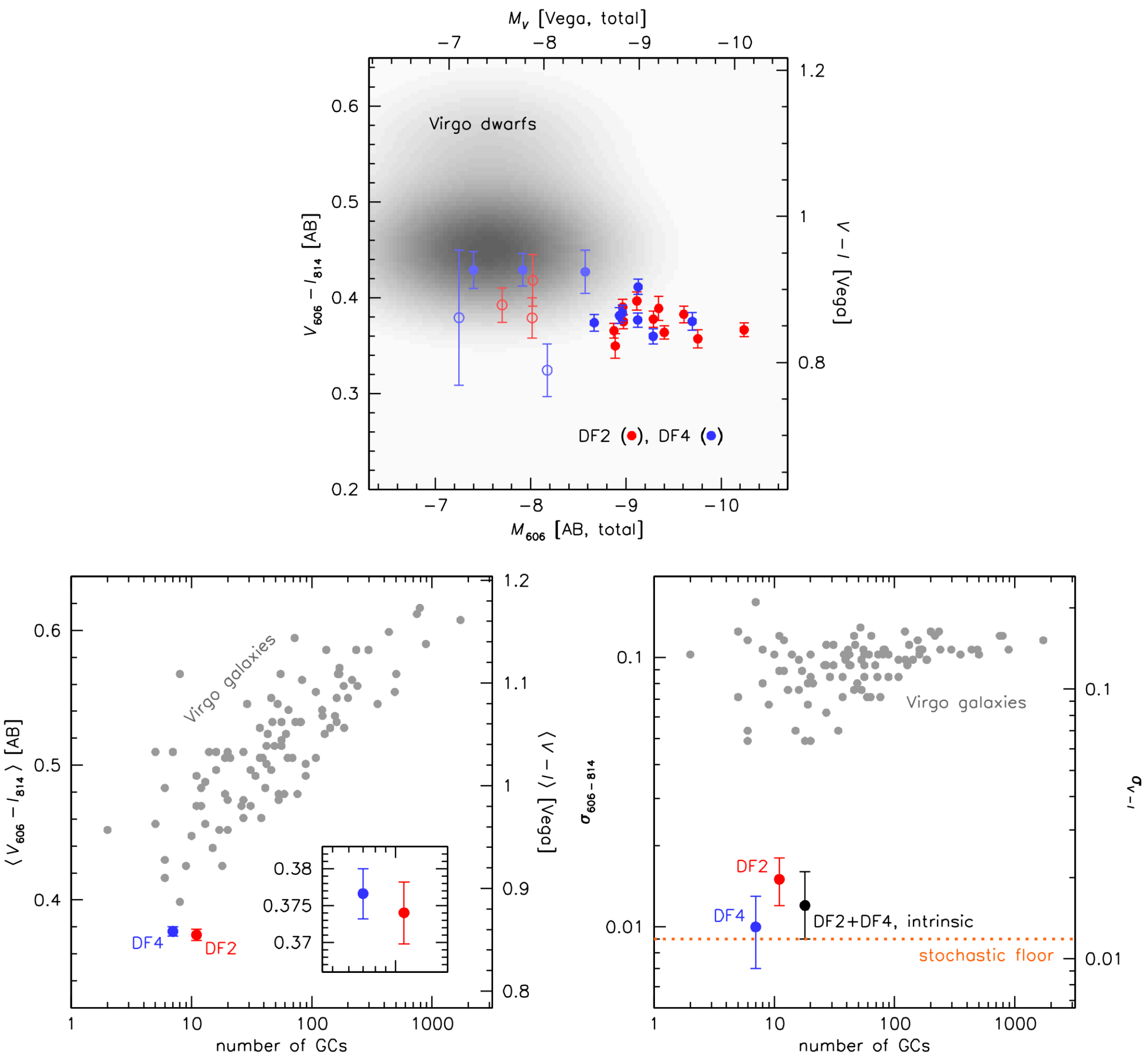}
  \end{center}
\vspace{-0.2cm}
    \caption{
{Top panel:} DF2/DF4 globular clusters in the color--magnitude plane. {Colors are corrected for Galactic reddening.} Solid symbols are
spectroscopically-confirmed and light-colored symbols have uncertainties $>0.015$\,mag.
The parameterized distribution for globular clusters
in low luminosity Virgo galaxies is shown in grey (see text).  
{Bottom left panel:} Mean {de-reddened} color of globular clusters with uncertainties $<0.015$, compared
to Virgo galaxies from \citet{peng:06}. There is no significant systematic color difference between DF2 and DF4.
{Bottom right panel:} Observed and intrinsic scatter
in the colors, again compared to Virgo galaxies. The clusters in DF2 and DF4 are extremely homogeneous.
The orange line is the minimum possible scatter, arising from stochastic
fluctuations in the number of red giants.
}
\label{cm1.fig}
\end{figure*}

\subsection{Aperture Photometry}

Fluxes are measured directly from the cleaned images using simple aperture photometry. Colors are measured
in $0\farcs 5$ diameter apertures and total $V_{606}$ magnitudes in $1\farcs 5$ diameter
apertures. The background is determined
from an annulus with an inner diameter of $1\farcs 5$ and an outer diameter of $3\farcs 0$.
As noted above the results depend on the precise choice of these parameters when applied to the
original images, but we find that they are insensitive to them when applied to the cleaned images.

Errors on the fluxes were determined empirically by repeating the aperture photometry in a grid of
$6\times 6$ blank positions within the $7\arcsec \times 7\arcsec$ cleaned images. This procedure ensures that the
local environment of each globular cluster is taken into account. As can be seen in the mask
images of Fig.\ \ref{figure1.fig}, the central region (where the globular cluster is) is typically not masked as
faint stars do not exceed the thresholds there. This leads to a bias that can be quantified and corrected
for in the grid photometry. At each grid position we replaced the pixels within $\pm 0\farcs 25$ of that
position with those from the original image.
The grid photometry now closely resembles the globular cluster photometry.
The mean flux from the 36 grid positions is subtracted from the globular cluster measurements
and the scatter among them is taken as the uncertainty. Both values were determined with the biweight estimator
\citep{beers:90} as it is insensitive to outliers.

Aperture corrections are taken from \citet{bohlin:16};
as the half-light radii of the globular
clusters are typically $0\farcs 05$ \citep{dokkum:18b}, these point source corrections are adequate
for obtaining accurate colors.
Fluxes are converted to AB magnitudes using the zeropoints in the headers of the images and
corrected for Galactic extinction using the \citet{schlafly:11} maps. The corrections are
$-0.060$\,mag, $-0.062$\,mag in $V_{606}$ for DF2, DF4,
and $-0.037$\,mag, $-0.038$\,mag in $I_{814}$.
Finally, total magnitudes are converted to absolute magnitudes.

\section{Observed color variation}
\label{obs.sec}

The distribution of the DF2 and DF4 globular clusters in the color--magnitude plane is shown in
the top panel of Fig.\ \ref{cm1.fig}. Errorbars are a combination of the measurement uncertainty
and the $0.007$\,mag flat fielding uncertainty. For reference,
the greyscale background shows the parameterized distribution of globular clusters in low luminosity
galaxies in the Virgo cluster. This is a combination of the two-component decomposition of the
color distribution for the faintest galaxies\footnote{Specifically, we averaged the fit values for the $M_B=-16.6$ and $M_B=-15.7$ bins in Table 4 of \citet{peng:06}. The $g-z$ colors were converted using $V_{606}-I_{814} = 0.445 (g-z) + 0.061$ {\citep[from Appendix A in][]{usher:12}.}}
in \citet{peng:06} and the Gaussian fit to the globular cluster
luminosity function of the faintest galaxies\footnote{We averaged the $g$-band fit values for the $M_B=-16.6$, $M_B=-16.4$, and $M_B=-15.7$ bins in Table 3 of \citet{jordan:07} and converted to $V_{606}$ using $V_{606}=g - 0.40$.}
in \citet{jordan:07}.

The globular clusters in DF2 and DF4 are much brighter than those in Virgo dwarfs, as has been discussed
extensively in earlier papers \citep[see][]{dokkum:18b,trujillo:19,shen:21a}, and they are also somewhat bluer.
There is no evidence for a systematic trend with magnitude when the full sample of confirmed and candidate clusters is considered. The photometric uncertainties increase sharply at fainter magnitudes, and
in the following we only consider the 18 clusters with errors $<0.015$. This sample corresponds to the
full sample of clusters with $M_{606}<-8.6$.  

The mean color of the DF2 and DF4 clusters is nearly identical: $\langle V_{606}-I_{814}\rangle=0.374\pm 0.004$
for DF2 and $\langle V_{606}-I_{814}\rangle= 0.377\pm 0.003$ for DF4,\footnote{The errorbars do
not include an uncertainty of $\approx 1$\%
in the absolute calibration of the ACS filters, as this systematic error affects all colors and magnitudes
by the same amount.} as determined
with the biweight estimator \citep{beers:90}.
The mean color difference is $\Delta_{\rm DF2-DF4}=-0.003 \pm 0.005$.
The mean colors are compared to those of globular clusters in Virgo galaxies\footnote{The
Virgo data are the single-component fits in Table 3 of \citet{peng:06}.}
in Fig.\ \ref{cm1.fig} (lower left).
The clusters
in DF2 and DF4 are bluer than those in Virgo galaxies with similar $N_{\rm GC}$.\footnote{The
color difference partly reflects an age difference: the DF2 clusters have ages of $9\pm 2$\,Gyr
\citep{dokkum:18b,fensch:19} whereas the typical
ages of metal-poor globular clusters are similar to those in the Milky Way \citep[see][]{strader:05}. {However, this only explains $\sim 1/3$ of the difference. We note that the colors of the DF2 and DF4 clusters are similar to those of clusters in the Milky Way: using colors from Table 2 of \citet{bellini:15} and metallicities from \citet{harris:96} (2010 edition) we find $\langle V_{606}-I_{814}\rangle_{\rm MW} \approx 0.38$ for [Fe/H]$=-1.3$.}}
Furthermore, in the Virgo
sample with $N_{\rm GC}<20$ the median color difference
between any two data points is $0.036$, an order of magnitude larger than the difference
between DF2 and DF4.

The cluster-to-cluster scatter is also very small, and within the errors is the same for the two galaxies:
$\sigma_{\rm obs}=0.015\pm 0.003$ for DF2 and $\sigma_{\rm obs} = 0.010\pm 0.003$ for DF4.
The observed scatter in the combined sample of 18 bright
globular clusters in DF2 and DF4 is $\sigma_{\rm obs}=0.015\pm 0.002$ (all determined with
the biweight estimator; the rms is also $0.015$).
The intrinsic scatter $\sigma_{\rm intr}$ can be determined by constructing the likelihood function,
\begin{equation}
\mathcal{L} = \prod_{i=1}^{i\leq 18}
\frac{1}{\sqrt{2\pi}\sigma_{\rm eff}}
\exp\left[-0.5\left( \frac{c_i-\mu}{\sigma_{\rm eff}}\right)^2\right],
\label{like.eq}
\end{equation}
with $c_i$ the colors of the individual clusters, $\mu$ the mean,
and $\sigma_{\rm eff}^2 = \sigma_{\rm intr}^2 + e_i^2$.
We find an intrinsic scatter of $\sigma_{\rm intr}=0.012_{-0.003}^{+0.004}$.
As shown in Fig.\ \ref{cm1.fig} (lower right) the typical 
scatter in Virgo galaxies is $\sigma \approx 0.1$\,mag, and there are no galaxies with $\sigma<0.05$\,mag.
We conclude that the globular clusters in DF2 and DF4 form a remarkably homogeneous population,
as predicted by the bullet dwarf collision model.\footnote{{The observed scatter for both galaxies is $\approx 0.03$\,mag when colors are measured directly from the original {\tt flc} files. The improvements in flat fielding and photometry led to a factor of $\approx 2$ lower observed scatter, with each contributing about equally.}} This level of homogeneity is
not observed in normal dwarf galaxies.

\begin{figure*}[htbp]
  \begin{center}
  \includegraphics[width=0.95\linewidth]{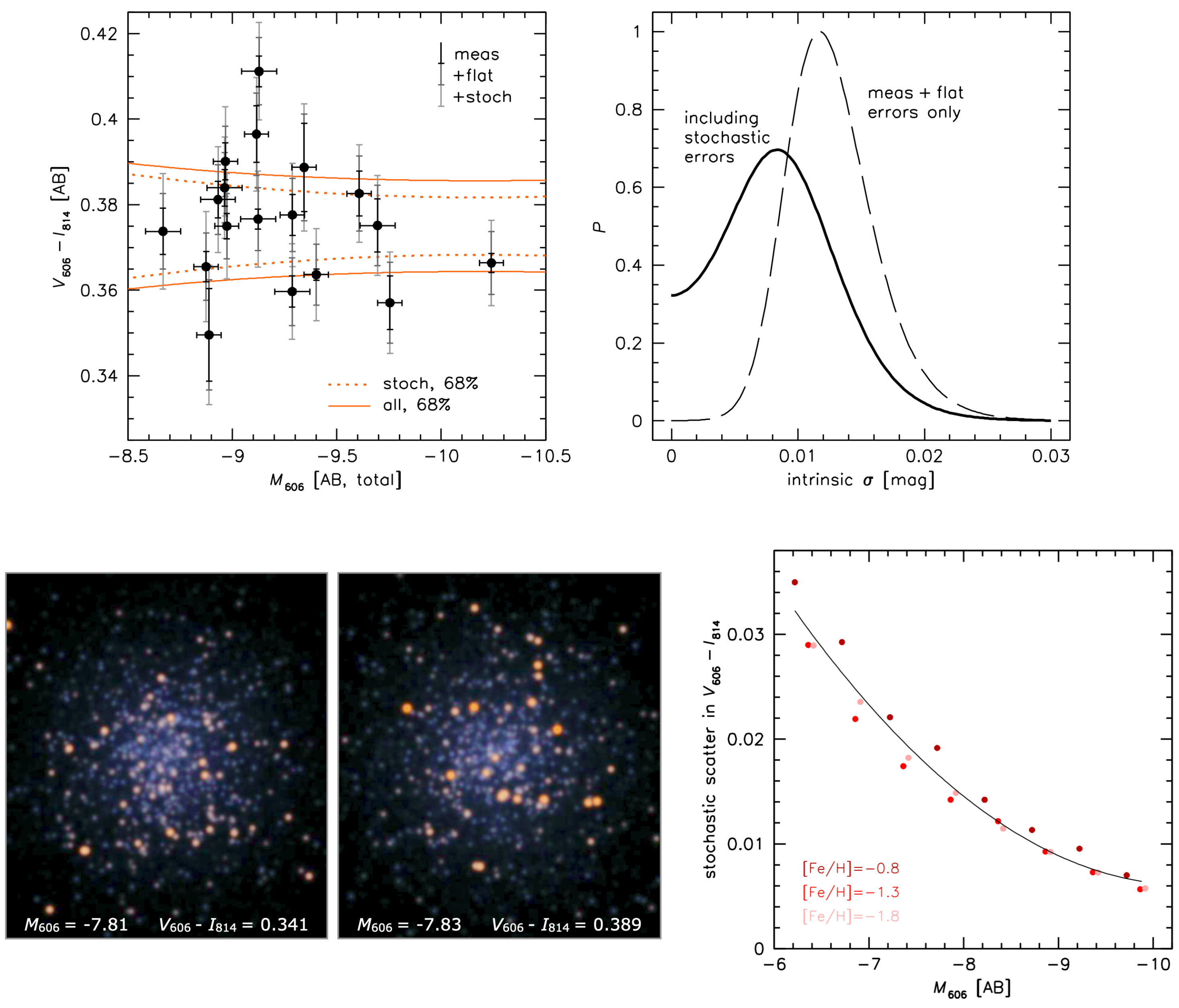}
  \end{center}
\vspace{-0.2cm}
    \caption{
{Top left panel:} Zoom-in on
color--magnitude measurements with errors $<0.015$, corresponding to all clusters with $M_{606}<-8.6$. {The colors are corrected for Galactic reddening.}
Errorbars are the measurement uncertainty (black), the combined measurement and flat-fielding uncertainty
(dark grey), and the combined measurement, flat-fielding, and stochastic sampling uncertainty (light grey).
Dotted orange lines show the expected $\pm 1\sigma$ variation due to stochastic sampling alone.
Solid lines include the other uncertainties. {Top right panel:} Likelihood for the intrinsic dispersion,
for the photometric errors (broken line) and for the total error, including the effect of stochastic sampling (solid line).
 {Bottom:} The effects of stochastic sampling
of the isochrone. Intrinsic
cluster-to-cluster color differences of $0.01-0.02$\,mag arise from random variations in the number of
red giants. Two examples of identical clusters with different random seeds are shown at left
(see text for details).
}
\label{cm2.fig}
\end{figure*}

\section{Interpretation of the variation}


The color variation is small but not zero. As noted above, observational uncertainties
explain some of the observed scatter.  This is shown in the top left
panel of Fig.\ \ref{cm2.fig}, where the errorbar for each data point is split into several distinct
contributions. Measurement uncertainties are shown in black and
the total photometric uncertainties, which include the $0.007\pm 0.003$\,mag flat fielding errors, are shown in
dark grey. 

The intrinsic scatter is so small that the effects of stochastic
sampling of the isochrone need to be taken into account. Red giants contribute significantly to the integrated
light and the Poisson variation in their number causes some clusters to be redder and brighter
than others. The same effect causes the well-known pixel-to-pixel surface brightness fluctuations in galaxy
images \citep{tonry:88,greco:21}. We quantify this effect by generating artificial globular clusters with the
{\tt ArtPop} code \citep{artpop} and measuring their integrated colors. The clusters have $\log({\rm age})=9.9$,
$-1.8 \leq [{\rm Fe}/{\rm H}] \leq -0.8$, and cover a factor of 20
in mass (parameterized by $\log(n_{\rm stars})$, which ranges from 5 to 6.3 with
steps of 0.2). At each mass 500 clusters are generated with different random seeds, and the rms scatter in the
$V_{606}-I_{814}$ colors is measured.

The results are shown in the bottom right panel of Fig.\ \ref{cm2.fig}. The cluster-to-cluster $V_{606}-I_{814}$
scatter is $\sim 1$\,\% in the relevant luminosity range, of the same order as the intrinsic scatter in the DF2/DF4
globular clusters, with a modest dependence on metallicity.
Two example clusters that illustrate the effect are shown at left
\citep[see also Fig.\ 6 in][]{artpop}. A polynomial fit 
to the luminosity-dependent variation has the form
\begin{equation}
\sigma_{\rm stoch}=0.019+0.0087(M_{606}+7.5)+0.00153(M_{606}+7.5)^2,
\end{equation}
and is shown by the black line in Fig.\ \ref{cm2.fig}.

Light grey errorbars in the top  panel of Fig.\ \ref{cm2.fig} show the effect of including this uncertainty
for each cluster. The broken orange lines show where 68\,\% of the points are expected to fall due to the
stochastic sampling effect alone. Solid orange lines include the measurement error; for a normal distribution entirely defined by these lines, 12 out
of 18 would fall within them. They encompass ten, only slightly fewer.
Using the total errors (that is, the combination of the measurement error, the
flat fielding error, and the stochastic variation) in the likelihood analysis gives
the ``stellar population scatter'', $\sigma_{\rm sp}
=0.008^{+0.005}_{-0.006}$. This scatter is not significantly different from zero.
The likelihood function, marginalized over $\mu$, is shown by the solid line in Fig.\ \ref{cm2.fig}.

\begin{figure*}[htbp]
  \begin{center}
  \includegraphics[width=0.7\linewidth]{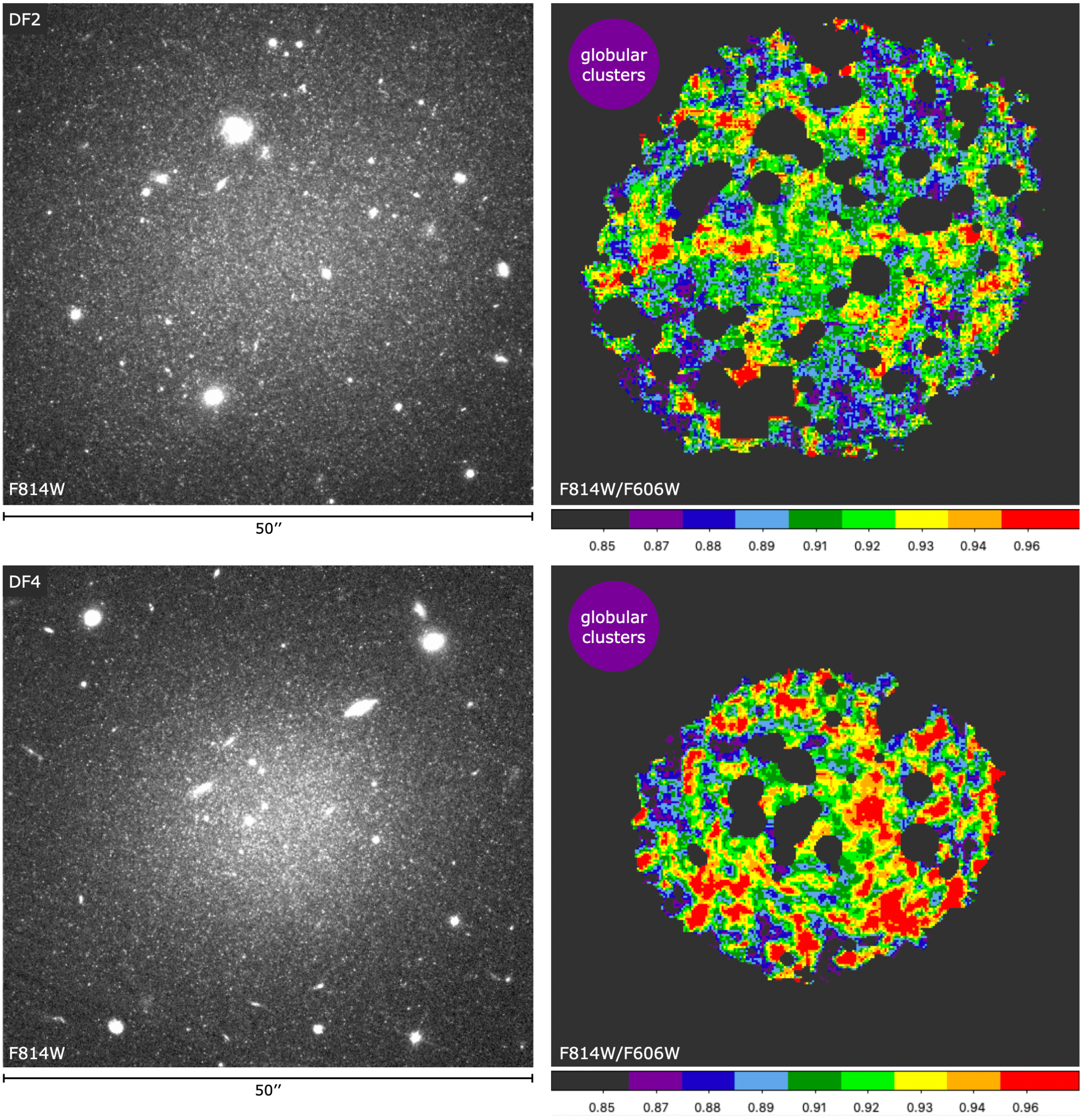}
  \end{center}
\vspace{-0.2cm}
    \caption{
Color images of the galaxies, created by dividing binned and median-filtered $I_{814}$ and $V_{606}$ images
of the galaxies. Numerical values are simply the ratios of the observed fluxes in e$^-$\,s$^{-1}$.
The mean value for the globular clusters is indicated with the circle.
Color variations within each galaxy are caused by surface brightness fluctuations. 
The average color of DF4 is slightly redder than that of DF2 but the difference is not significant:
$(V_{606}-I_{814})_{\rm DF2} = 0.420\pm 0.024$ and $(V_{606}-I_{814})_{\rm DF4}=0.436\pm 0.075$.
The galaxies are redder than the globular clusters.
}
\label{colorimages.fig}
\end{figure*}

\label{ssp.sec}

We use the flexible stellar population synthesis (FSPS) framework \citep{conroy:09}
with the MIST isochrones \citep{choi:16} to determine the 
(limits on) variation in age and metallicity that is
implied by the stellar population scatter. For ages near 8\,Gyr and metallicities near ${\rm [Fe/H]}=-1.3$ {\citep[as independently measured from stacked spectra by][]{dokkum:18b,fensch:19}}
we find 
$\Delta(V_{606}-I_{814})=0.10\Delta{\rm [Fe/H]}$ and
$\Delta(V_{606}-I_{814})=0.0064\Delta({\rm age})$ to good approximation, with age in Gyr.\footnote{{The slope of the color-metallicity relation is identical for Milky Way clusters: using data from Table 2 of \citet{bellini:15} and from \citet{harris:96} (2010 edition) we find $V_{606}-I_{814} = 0.10 {\rm [Fe/H]} + 0.50$ with an rms of 0.04\,mag.}}
The observed stellar population scatter $\sigma_{\rm sp}$ implies $\sigma_{\rm [Fe/H]} = 0.08^{+0.05}_{-0.06}$
if there is no variation in age
or $\sigma_{\rm age} = 1.3^{+0.8}_{-1.0}$\,Gyr if there is no variation in metallicity. 
In the simulations of \citet{lee:21} the globular clusters can have a spread
of up to $\approx 0.1$\,dex in metallicity and $\approx 150$\,Myr in age, and
we conclude that the bullet hypothesis cannot be ruled out.

\section{Colors of the Diffuse Light}

The bullet model also makes predictions for the global colors of DF2 and DF4, although these are
more model-dependent than the predictions for the globular clusters
\citep{shin:20,lee:21}. The galaxies are predicted to have higher metallicities than the globular
clusters as they form stars over a longer time period
\citep[see][]{lee:21}. Therefore, while the colors of DF2 and DF4
should be very similar to each other, they are predicted to be
redder than those of the globular clusters.

We measure the average colors of the galaxies in the following way. An object mask is created by
comparing the summed $V_{606}+I_{814}$ image to a binned and median-filtered version of itself. 
A first model for the galaxy is made by median filtering the $V_{606}$ and $I_{814}$ images, not taking masked pixels
into account. This model is subtracted from the data and the object mask is optimized with a lower
threshold, taking care not to include giants in the mask. Then the median filtering is repeated to create
a final model in each filter. There is a background gradient in all images, and at this stage a surface
is fitted to the background and subtracted. The rms of this background model (that is, 68\,\% of the gradient
that is removed) is taken as the uncertainty in each filter.
Finally, the $I_{814}$ model is divided by the $V_{606}$ model, multiplied by the object mask, and a
flux threshold is applied so that the faint outskirts are excluded.

These color images are shown in Fig.\ \ref{colorimages.fig}, and compared to the mean color of the
globular clusters. The variations within each galaxy are caused
by stochastic variation in the number of red giants (surface brightness fluctuations). We measure the
mean colors of the galaxies to be
$(V_{606}-I_{814})_{\rm DF2} = 0.420\pm 0.024$ and $(V_{606}-I_{814})_{\rm DF4}=0.436\pm 0.075$,
consistent with the \citet{cohen:18} measurements with smaller uncertainties.
We infer that  the colors of the galaxies are identical within the errors,
as predicted by the bullet model.

The galaxies are redder than the luminous globular clusters. If the actual color
of both galaxies is $V_{606}-I_{814}=0.375$ and the errors are Gaussian,
the probability that we measure $(V_{606}-I_{814})_{\rm DF2}\geq 0.420$
and $(V_{606}-I_{814})_{\rm DF4}\geq 0.436$ is $<1$\,\%. 
The color difference between the galaxies and the clusters
of 0.05\,mag is qualitatively consistent with the metallicity difference predicted
by the hydrodynamical model of \citet{lee:21}.
Using the relation in \S\,\ref{ssp.sec} we find $\Delta{\rm [Fe/H]}
\approx 0.5$, somewhat larger than the $\approx 0.2$ predicted by
\citet{lee:21} but in good agreement with the spectroscopically-determined
value of $\Delta{\rm [Fe/H]}=0.56\pm 0.15$ of \citet{fensch:19}.

\section{Discussion}

The central result of this paper is that the bright globular clusters in DF2 and DF4 have extremely similar
colors. The observed scatter is $\approx 0.015$\,mag, and this can be explained by a combination of
measurement uncertainties ($\approx 0.01$) and stochastic variations in the number of red giants
($\approx 0.01$). The remaining scatter among the 18 luminous clusters in DF2 and DF4
is $\sigma_{\rm ssp}=0.008^{+0.005}_{-0.006}$, that is, not significantly different
from zero. 
The diffuse light is
redder in both galaxies, with again no significant difference between DF2 and DF4. These results are
expected in the bullet dwarf scenario \citep[see][]{silk:19,shin:20,lee:21,dokkum:22} and we conclude that
this model survives an important falsification test. Returning to the formation models listed in \S\,1,
no other published explanation for the lack of dark matter in DF2 and DF4
also ``naturally'' produces the extreme uniformity of their globular cluster populations.

The globular clusters in DF2 and DF4 are different from those in Virgo galaxies; as detailed in
\S\,4 they are brighter, bluer, and have a much smaller scatter. DF2 and DF4 are ultra-diffuse galaxies
\citep{dokkum:15udgs} and at least two other
UDGs also have very homogeneous globular cluster populations, NGC\,5846-UDG1 \citep{muller:21,danieli:22}
and DGSAT\,{\sc i} \citep{janssens:22}. In other respects they are different; NGC\,5846-UDG1 has a canonical
globular cluster luminosity function, and DGSAT\,{\sc i}'s clusters are significantly redder
than those in DF2/DF4. A small scatter 
indicates synchronized formation in a dense, homogeneous medium and it may
arise generically in any formation scenario that results in the extreme (factor of $10^9$)
density contrast between the globular clusters and the galaxy light that is seen in UDGs
\citep[see, e.g.,][]{dokkum:18b,trujillogomez:21}. Specifically, intense feedback from
the formation of the globular clusters may have caused the galaxies to expand and turn into
UDGs \citep{trujillogomez:21,danieli:22}. Further deep HST or JWST
studies of globular clusters in UDGs in various environments are needed to investigate this further.

Our analysis focuses on the brightest clusters as these have the smallest uncertainties. \citet{shen:21a}
showed that the
globular cluster luminosity function in DF2 and DF4 can be modeled as a combination of a bright peak
of overluminous clusters
plus a ``normal'' luminosity function with a normalization and mean that are typical for the galaxies'
luminosities. In this context it is interesting that the clusters with $M_{606}>-8.6$ are somewhat redder
than the brighter ones, with $\langle V_{606}-I_{814}\rangle=0.400\pm 0.011$.
Perhaps the faint clusters formed together with the diffuse light, with a higher mean metallicity
than the brighter ones.\footnote{\cite{lee:21} predict the opposite trend, with the brighter clusters having
a higher metallicty. {We also note that the Lee et al.\ simulations produce a much larger number of low mass clusters than are observed in DF2 and DF4.}}
The observed scatter is also higher for the faint clusters
at $\sigma_{606-814}=0.034 \pm 0.008$, although this can be explained by
a combination of measurement errors ($0.028$) and stochastic sampling ($0.016$).
Extremely deep spectroscopy may shed more might on these questions.

Further tests of the bullet model are possible. The most straightforward next step is probably obtaining
radial velocities and line-of-sight
distances of other galaxies along the trail, as the bullet model predicts that
these follow a regular sequence (with some contamination from
unrelated objects).  Dynamical mass measurements of other trail galaxies will also
be highly constraining, and interesting in a broader context as
bullet dwarf events can, in principle, constrain the self-interaction cross section of dark matter. Modeling of
the bullet cluster has provided an upper limit \citep{randall:08}, but as self-interacting dark matter was introduced to explain the ``cored'' dark matter density profiles of low mass galaxies \citep{spergel:00} it is important to measure the cross section on those scales \citep{tulin:18}.

\vspace{0.5cm}
\noindent
We thank Yotam Cohen and the rest of the STScI ACS team for their
help with the data reduction, and the anonymous referee for their insightful comments and
suggestions.
Support from HST grants GO-14644, GO-15695, and GO-15851 is gratefully acknowledged.
S.~D.\ is supported by NASA through Hubble Fellowship grant HST-HF2-51454.001-A.


\bibliography{cm}{}
\bibliographystyle{aasjournal}

\end{document}